\documentstyle[preprint,cite,aps]{revtex}
\begin{document}
\draft
\title{\Large On the equivalence of the self-dual and 
Maxwell-Chern-Simons models coupled to Fermions}
\author{M. Gomes and L. C. Malacarne\footnote{Permanent address:
    Departamento de F\'\i sica, Universidade Estadual de Maring\'a -
    Av. Colombo, 5790 - 87020-900, Maring\'a -PR,Brazil} and A. J. da Silva}
  \address{Instituto de F\'\i sica, USP\\
 C.P. 66318 - 05315-970, S\~ao Paulo - SP, Brazil}
\date{1997}

\maketitle

\begin{abstract}
We study the exact equivalence between the self-dual model minimally coupled
with a Dirac field and the Maxwell-Chern-Simons model with non-minimal
magnetic coupling to fermions. We show that the fermion sectors of the
models are equivalent only if a Thirring like interaction is
included. Using functional methods we verify that, up to
renormalizations, the equivalence persists at the quantum level.

\end{abstract}


\newpage
(2+1) dimensional models have provided insights into some
basic aspects of field theory as the existence of massive gauge
fields and anomalous spin and statistics. In this context, it is important 
to establish connections between  apparently unrelated
models so that  unifying pictures could emerge. Examples of such
endeavor are the recent tentatives of bosonization, trying to blend
fermions and bosons within a coherent framework. Some approaches
to bosonization rest on the fact that 
in (2+1) dimensions there are two ways to describe a single, freely
propagating, spin one, massive mode. In fact, one can use either the
gauge non-invariant self-dual model ($SD$), originally proposed by
Townsend et al.\cite{Thousend}, or the topologically massive gauge
invariant Maxwell-Chern-Simons model($MCS$)\cite{Jackiw1}.  These two
descriptions are actually equivalent and this equivalence holds even
if the basic fields are coupled to external sources \cite{Jackiw2}.
Additional studies on the connection of these models,
[$\citen{Gianvittorio}$-$\citen{Banerjee2}$] have shown that the SD is a
gauge fixed version of the MCS model.

The mentioned relation between the SD and the MCS models has been used
to map the massive Thirring model into a theory of interacting
bosons\cite{Fradkin}. More precisely, to leading order in the inverse
fermionic mass, the use of the correspondence allows to identify the
Green functions of the currents of the massive Thirring model with
those of the basic vector field of th MCS
model.  The extension of these results to the non Abelian situation
has been considered by various authors
\cite{Karlhede,Arias,Bralic,Guillou,Banerjee3}.  

Although the
equivalence of the SD and MCS models is well established in the free field case,
it remains to understand what happens when the coupling to other
dynamical fields is considered.  This letter is specifically devoted
to the study the equivalence when the SD and MCS are coupled to Dirac
matter fields. By inspection, the source current for the SD field is found to
be
replaced by its  rotational in the MCS vector field equation. Moreover,
as we shall see, in opposition to what happens in the external source
case, the classical equivalence of the corresponding
Dirac's equations already demands the inclusion of an additional Thirring like
interaction.
At the
quantum level one must be more careful because  some of the interactions
are not perturbatively renormalizable. 

We begin our analysis by recalling that, at the field equations level,
the SD model described by
\begin{eqnarray}
\label{1}
{\cal L}_{SD} = \frac 12 m^2 f_\mu f^\mu - \frac m2 
\varepsilon ^{\mu\nu\rho} f_\mu\partial_\nu f_\rho \; ,
\end{eqnarray}
is connected to the $MCS$ model,  
\begin{eqnarray}
\label{2}
{\cal L}_{MCS} = -\frac 14 F_{\mu\nu} F^{\mu\nu} + \frac 14 m
 \varepsilon ^{\mu\nu\rho} A_\mu F_{\nu\rho}\; , 
\end{eqnarray}
where $ F_{\mu\nu}= \partial_\mu A_\nu - \partial_\nu A_\mu$, through the identification 
$f^\mu \leftrightarrow F^\mu = \frac{1}{ m} \varepsilon ^{\mu\nu\rho}
\partial_\nu A_\rho$, i. e., the basic field of the SD model corresponds to
the dual of the MCS field. 

Consider now a Dirac field minimally coupled to a vector field 
specified by the SD model, so that the new Lagrangian becomes

\begin{eqnarray}
\label{3}
{\cal L}_{SD}^{min} = {\cal L}_{SD}  - e f_\mu J^\mu + 
\bar{\psi}(i\partial\!\!\! /  -M)\psi \; , 
\end{eqnarray}
where $J^\mu = \bar{\psi}\gamma^\mu \psi$ and $M$ is the fermion
mass\cite{notacao}.  From the equivalence between the SD and MCS
models and noticing that the same identification of the fields, namely
$f^\mu \leftrightarrow F^\mu$, up a total derivative, changes the
minimal coupling into a magnetic coupling , i. e.,
\begin{equation}
f_\mu J^\mu \rightarrow  A_\mu G^\mu,
\end{equation}
where $ G^\mu= \frac 1m \varepsilon ^{\alpha\beta\mu} \partial_\alpha
J_\beta$, we may infer the equivalence of the model (\ref{3}) with the
one specified by
\begin{eqnarray}
\label{4}
{\cal L}_{MCS}^{mag} = {\cal L}_{MCS}  - e A_\mu G^\mu + 
\bar{\psi}(i\partial\!\!\! / -M)\psi \; . 
\end{eqnarray}

Indeed, a direct  inspection of the equations of
motion for the fields $f^\mu$ and $A^\mu$,
\begin{eqnarray}
\label{5}
-m\varepsilon^{\nu\alpha\beta}\partial_\alpha f_\beta + m^2 f^\nu &=& e
J^\nu \\
-m\varepsilon^{\nu\alpha\beta}\partial_\alpha F_\beta + m^2 F^\nu &=& e
G^\nu\;\label{5a} .
\end{eqnarray}
shows that the models are classically equivalents if the
identification
\begin{eqnarray}
\label{6}
f^\mu \leftrightarrow F^\mu \Longrightarrow J^\mu \leftrightarrow
G^\mu \; ,
\end{eqnarray}
is made. This indicates that the electric interaction goes into a magnetic one.
However, we still have  to examine Dirac's equation. We have,
\begin{equation}
(i\not \! \partial - M ) \psi= e f_\mu \gamma^\mu \psi,
\end{equation}
in the case of the SD model, and
\begin{equation}\label{6a}
(i\not \! \partial - M ) \psi= -\frac{e}{m} \epsilon^{\alpha\beta \rho}
\partial_\alpha A_\rho \gamma_\beta \psi
\end{equation}
for the MCS.  From (\ref{5a}) we can express the gauge field $A_\mu$ in terms
of the source $G^\mu$. We get,
\begin{equation}
A_\alpha(x) = e \int dy D_{R\mu \alpha}(x-y) G^{\mu}(y)\label{6b}
\end{equation}
where, in momentum space,
\begin{equation}
D_{R\mu \alpha}(k)= \frac{1}{k^2-m^2+ i\epsilon k_0}\left[ g_{\mu\alpha}
- \frac{k_\mu k_\alpha}{k^2+ i \epsilon k_0 }-i m
\varepsilon_{\mu\alpha\rho}\frac{k^\rho}{k^2+ i \epsilon k_0}\right ]
+ \mbox{gauge fixing term}
\end{equation}
is the retarded propagator for the self dual field. Replacing (\ref{6b}) into 
(\ref{6a}) one obtains
\begin{equation}
(i\not \! \partial - M ) \psi= e f_\mu \gamma^\mu \psi +\frac{e^2}{m^2} 
J^\beta\gamma_\beta \psi
\end{equation}
\noindent
where  
\begin{equation}
f_\mu(x)= e \int dy\, \Delta_{R\mu\alpha}(x-y) J^\alpha (y)
\end{equation}
and
\begin{equation}
\Delta_{R\mu\alpha}(k)= \frac{1}{k^2-m^2+i \epsilon k_0}\left[ g_{\mu\alpha}
- \frac{k_\mu k_\alpha}{m^2 }-i m
\varepsilon_{\mu\alpha\rho}\frac{k^\rho}{m^2}\right ] \label{7}
\end{equation}
is the retarded propagator for the MCS field. Basic to this result
is the algebraic identity
\begin{equation}\label{7a}
-\frac{i}{m}\epsilon^{\alpha\beta \mu}k_\alpha D_{R\mu\nu}(k) \frac{i}{m}
\epsilon^{\sigma\rho\nu}k_\sigma= \Delta_{R}^{\beta\rho}(k)+ \frac{g^{\beta
\rho}}{m^2} 
\end{equation}
relating the SD and MCS propagators.

Summing up,  our result
shows  that the fermionic sectors of the two models will agree only
if one includes in one of the models a Thirring like  interaction.
It should be clear that the Thirring interaction could be either incorporated
to the SD model, as we did, or to the MCS model, as we will do shortly.

We now want to investigate to what extension the equivalence (\ref{6})
holds at the quantum level. Here we expect some problems as the system
described by (\ref{3}) is perturbatively renormalizable, whereas the
Lagrangian (\ref{4}) is non-renormalizable. This can be seen by direct
power counting. The mass dimensions of the $A_\mu$ and $\psi$ are
$1/2$ and $1$, respectively. Therefore the mass dimensions of the
magnetic and of the Thirring  interactions are equal to $7/2$ and $4$, respectively,spoiling the renormalizability
of the models.  However this difficulty can be eventually surmounted.
For example, if there are $N$ fermions the theories turns out to be
$1/N$ expandable and in this context they are renormalizable.
For a solid consideration, we will employ the generating
functional
\begin{equation}
Z(\psi)= \int Df^\mu DA^\nu \exp[i\int {\cal L}(f,A,\psi)]\, ,
\end{equation}
using a master 
Lagrangian  which generalizes (\ref{3}) and (\ref{4}),
\begin{equation}
{\cal L}= \frac{m^2}{2} f_\mu f^\mu - m^2 f_\mu F^\mu
+ \frac{m^2}{2} F^\mu A_\mu -e J_\mu f^\mu 
-e A^\mu g_\mu+ \frac {\lambda}{2} (\partial_\mu A^\mu)^2
+ {\cal L}_D,
\end{equation}
where ${\cal L}_D$ is the free Dirac Lagrangian and $g_\mu$, $J_\mu$ are
 matter currents depending only on the fields $\psi$ and $\bar \psi$.
We take $J_\mu$ as the current $\bar \psi \gamma_\mu \psi$ but leave
the specific form of $g_\mu$ unspecified for the time being.
Integrating over the field configurations $f_\mu$ gives
\begin{equation}
Z(\psi)= \int DA^\mu \exp[i\int {\cal L}^{(1)}_{eff}(A,\psi)],
\end{equation}
where
\begin{eqnarray}
{\cal L}^{(1)}_{eff}(A,\psi) &=& -\frac 14 F_{\mu\nu} F^{\mu\nu} + \frac 14 m
 \varepsilon ^{\mu\nu\rho} A_\mu F_{\nu\rho} + \frac \lambda 2 (\partial_\mu
A^\mu)^2 -\nonumber \\
&&e  A^\mu( g_\mu+G_\mu ) -\frac {e^2}{2m^2} J^\mu J_\mu
+{\cal L}_D \, .\label{8}
\end{eqnarray}

On the other hand, integrating over $A_\mu$ furnishes
\begin{equation}
Z(\psi)= \int Df^\mu \exp[i\int {\cal L}^{(2)}_{eff}(f,\psi)],
\end{equation}
where
\begin{eqnarray}
{\cal L}^{(2)}_{eff}(f,\psi) &=& \frac {m^2}2 f^\mu f_\mu - \frac m{2} 
\varepsilon_{\alpha \beta\gamma} f^\alpha \partial^\beta f^\gamma 
-e\, f^\alpha[J_\alpha+ (g_{\alpha\beta} - \frac{\partial_\alpha
\partial_\beta}{\partial^2}) g^\beta]+ \nonumber \\
&& \frac {e^2}{2m} g_\mu[\varepsilon^{\mu\beta\nu} 
\frac{\partial_\beta}{\partial^2}+ \frac{1}{\lambda}
\frac{\partial^\mu\partial^\nu}{(\partial^2)^2}] g_\nu 
+ {\cal L}_D\, .\label{9}
\end{eqnarray}

We may now consider  some possibilities:

1. $g_\mu=0$.  In this situation, (\ref{8}) and (\ref{9}) prove the quantum
equivalence of the  models previously discussed.

2. If instead we take  $g_\mu= - G_\mu$ then the MCS field decouples whereas
the fermions  interacts through the  Thirring interaction.  This model of noninteracting fermions
and  vector fields is equivalent to the model in which, besides
a self interaction, the fermions are coupled through a self dual field. This
later model is described by the Lagrangian 

\begin{eqnarray}
{\cal L}^{(2)}_{eff}(f,\psi) &=& \frac {m^2}2 f^\mu f_\mu - \frac m{2} 
\varepsilon_{\alpha \beta\gamma} f^\alpha \partial^\beta f^\gamma 
-e f^\alpha (J_\alpha-G_\alpha)-\frac {e^2}{2m^3}\varepsilon_{\alpha \beta\gamma}
J^\alpha\partial^\beta J^\gamma   
\end{eqnarray}

3. As a last possibility we take $g^\mu=-J^\mu$. As it happens (\ref{9}) becomes
\begin{equation}
{\cal L}^{(2)}_{eff}={\cal L}_{SD} + {\cal L}_{D}+\frac{e^2}{2 m}
\varepsilon^{\mu\beta\nu}J_\mu\frac{\partial_\beta}{\partial^2} J_\nu,
\end{equation}
describing free self dual vector fields and self interacting fermions.
By (\ref{8}) this model is then equivalent to the one in which fermions
and vector fields interact as specified by the effective Lagrangian
\begin{equation}
{\cal L}^{(1)}_{eff}={\cal L}_{MCS} +{\cal L}_D +e A^\mu(J_\mu-G_\mu)
-\frac{e^2}{2 m^2}J_\mu J^\mu + \frac \lambda2 (\partial_\mu A^\mu)^2
\end{equation}

It should be clear that the equivalence of models that we just proved is
very particular, being a direct consequence of an algebraic identity. For
example it does not seem to hold in the scalar case. Indeed, in that
case the gauge current depends explicitly on the gauge field and, accordingly,
the functional determinant  has a non trivial dependence on the
matter fields.                  

The models considered involve perturbatively non renormalizable couplings,
namely the Thirring and the magnetic interactions. However, if the
fermion field is an $N$ component field the models are $1/N$ expandable and
in this context they are renormalizable. For the Thirring model this was 
proved in \cite{Go}. Here we examine the case when the MCS field
interacts  through the magnetic coupling, $A_\mu G^\mu$.
It is easy to verify that the propagator for the vector field behaves as the 
momentum $p$ tends to infinity as $1/(p^2)^{3/2}$. Power counting, therefore, establishes   that the
degree of superficial divergence of a generic graph  $G$ is
\begin{equation}
d(G) = 3 -N_F \label{10}
\end{equation} 
where $N_F$ is the number of external fermion lines of $G$. Apparently,
here we run into a difficulty since $d(G)$ does not decreases as the
number of external vector lines  increases. However, (\ref{10}) does not take into account that,  for an external vector line, the  momentum factor
associated with the vertex to which the line is attached does not
depend on the integration variables so that the effective degree  of divergence is lowered to
\begin{equation}
 3 -N_F-N_B\, ,
\end{equation}
which, as expected, is the same as in the Thirring model. It is also
easy to
verify that the counterterms needed to render finite the Green functions
have the same form of terms already present in the original Lagrangian.
This means that the theory is renormalizable as, stated.

\section*{acknowledgments}

This  work was partially supported by Conselho Nacional de Desenvolvimento
Cient\'{\i}fico e Tecnol\'ogico (CNPq) and Coordena\c c\~ao de Aperfei\c coamento de Pessoal de N\'{\i}vel Superior (Capes).


\begin{thebibliography}{99}
\bibitem{notacao} We use natural units ($c=\hbar=1$) and our metric is
  $g_{00}=-g_{11}=-g_{22}=1$. The fully antsymetric tensor
  $\varepsilon^{\mu\nu\rho}$ is normalized such that
  $\varepsilon^{012}=1$. For the $\gamma$-matrices we adopt the
  representation $\gamma^0=\sigma^3$, $\gamma^1=\sigma^1$,
  $\gamma^2=i\sigma^2$, where $\sigma^i$, $i=1,2,3$, are the Pauli
  spin matrices.
\end{thebibliography}
\end{document}